\documentclass[aps,prc,twocolumn,amsmath,superscriptaddress,nofootinbib]{revtex4-1}

\usepackage{varwidth}
\usepackage{multirow}
\usepackage{dcolumn}
\usepackage{tabularx}
\usepackage{booktabs}
\newcolumntype{C}{>{\centering\arraybackslash}X}

\usepackage{graphicx}
\usepackage{epstopdf}
\usepackage{graphics}
\usepackage{graphicx}
\usepackage{epsfig}
\usepackage{epstopdf}
\usepackage{makecell}
\usepackage{hyperref}
\hypersetup{
    colorlinks=true,
    linkcolor=blue,
    citecolor=blue,
    urlcolor=blue,
    anchorcolor=blue}

\begin{document}
\title{Identifying an $\rm X(3872)$ tetraquark state versus a molecular state by
         formation time, velocity and temperature in relativistic nuclear collisions}

\author{Zhi-Lei She}
\email[]{shezhilei@cug.edu.cn}
\affiliation{School of Mathematical and Physical Sciences, Wuhan Textile
            University, Wuhan 430200, China}

\author{An-Ke Lei}
\affiliation{Key Laboratory of Quark and Lepton Physics (MOE) and Institute of
            Particle Physics, Central China Normal University, Wuhan 430079,
            China}

\author{Yu-Liang Yan}
\affiliation{China Institute of Atomic Energy, P. O. Box 275 (10), Beijing
            102413, China}

\author{Dai-Mei Zhou}
\email[]{zhoudm@mail.ccnu.edu.cn}
\affiliation{Key Laboratory of Quark and Lepton Physics (MOE) and Institute of
            Particle Physics, Central China Normal University, Wuhan 430079,
            China}

\author{Wen-Chao Zhang}
\affiliation{School of Physics and Information Technology, Shaanxi Normal
University, Xi'an 710119, China}

\author{Hua Zheng}
\affiliation{School of Physics and Information Technology, Shaanxi Normal
University, Xi'an 710119, China}

\author{Liang Zheng}
\affiliation{School of Mathematics and Physics, China University of Geosciences
            (Wuhan), Wuhan 430074, China}

\author{Yi-Long Xie}
\affiliation{School of Mathematics and Physics, China University of Geosciences
            (Wuhan), Wuhan 430074, China}

\author{Gang Chen}
\affiliation{School of Mathematics and Physics, China University of Geosciences
            (Wuhan), Wuhan 430074, China}

\author{Ben-Hao Sa}
\email[]{sabhliuym35@qq.con}
\affiliation{Key Laboratory of Quark and Lepton Physics (MOE) and Institute of
            Particle Physics, Central China Normal University, Wuhan 430079,
            China}
\affiliation{China Institute of Atomic Energy, P. O. Box 275 (10), Beijing
            102413, China}

\date{\today}

\begin{abstract}
The production of exotic hadron $\rm X(3872)$ in $pp$ collisions at
$\sqrt{s}=2.76\,\mathrm{TeV}$ is investigated by the parton and hadron
cascade model PACIAE in this work. In the simulation the final
partonic state (quark matter, QM) and the final hadronic state (hadron matter,
HM) are continuously processed and recorded. The $\rm X(3872)$ compact
tetraquark state and loose molecular state are, respectively, coalesced and
recombined in the QM and HM with the quantum statistical mechanics inspired
dynamically constrained phase-space coalescence model. The
formation time, velocity and temperature of QM (tetraquark state) and HM (molecular
state) are proposed as identifying criteria between the two states. Our results in
transverse momentum spectrum and rapidity distribution, etc. show a
significant discrepancy between the two states and confirm that they are also valuable
criteria identifying the $\rm X(3872)$ compact tetraquark state or molecular
state.
\end{abstract}

\maketitle

\section{Introduction}
Exotic states such as unconventional multiquark configurations containing more
than three valence quarks are allowed and expected by the quark model and the
quantum chromodynamics (QCD)~\cite{gell1964,jaffe1977}. The $\rm X(3872)$ as the first exotic candidate
was observed in $e^{+}e^{-}$ collisions by the Belle Collaboration in 2003~\cite{belle2003}.
Later, more experiments have confirmed its existence and have measured its
properties in both $pp$ collisions~\cite{lhcb2013,cms2013,atlas2017,lhcb2021,lhcb2024dj}
and nucleus-nucleus collisions~\cite{cms2022}.

A key issue related to the $\rm X(3872)$ production is its internal structure in
relativistic nuclear collisions~\cite{cho2017}. Various interpretations of the $\rm X(3872)$ have been proposed and studied,
such as the hadronic molecular state of $D \bar D^{*}$~\cite{braaten2004,tornqvist2004}, compact multiquark
state~\cite{maiani2005,matheus2007}, hybrid meson~\cite{close2003,li2005},
and charmonium~\cite{eichten2004,suzuki2005}, etc. However, no consensus has been reached yet.
The latest theoretical analysis indicates that the $\rm X(3872)$ is likely to be a tetraquark
state~\cite{esposito2021,tai2023} or a molecular state~\cite{yun2023}.
We refer to recent reviews~\cite{chen2016,esposito2017,guo2018,olsen2018,nora2020,chen2023} for more
details.

Several observables are proposed to distinguish the $\rm X(3872)$
molecular state or tetraquark state: The yield ratio of coalescence model
to the statistical model~\cite{cho2011} was recommended. Then the spatial
structure parameter ($R_{0}$) was followed~\cite{chen2016,ge2021,hui2021}.
Furthermore, the rapidity distribution, transverse momentum spectrum,
and elliptic flow ($v_2$) were suggested~\cite{hui2021}. Subsequently, the yield centrality
dependence was proposed~\cite{wu2021,bao2022}. More recently, the radiative
decay ratio has also been suggested~\cite{esposito2021,grinstein2024}.

The parton and hadron cascade model PACIAE~\cite{lei2023}
together with the dynamically constrained phase space coalescence
model (DCPC~\cite{yan2012}) has been successfully applied
describing the production of light nuclei, such as deuteron and hypertriton~\cite{yan2012,she2021}, and
the exotic hadrons, such as $\mathrm X(3872)$, $P_{c}$ states, and
$\chi_{c1}(3872)/\psi(2S)$ cross-section ratio, in
relativistic nuclear collisions~\cite{tai2023,ge2021,hui2022}.
However, in the previous works, the exotic hadron is coalesced in the hadronic final state,
despite the compact tetraquark state or loose molecular state, the same as Ref.~\cite{zzhen2021}.

In this paper, the PACIAE + DCPC model is used to simulate the
production of $\rm X(3872)$ in $pp$ collisions at $\sqrt{s}=2.76\,\mathrm{TeV}$.
The $\rm X(3872)$ tetraquark state and
the molecular state are coalesced and recombined, respectively, by the DCPC model
in the final partonic state (FPS)
and the final hadronic state (FHS), which are simulated by PACIAE model.
The formation time, velocity and temperature of
FPS (tetraquark state) and FHS (molecular state) are proposed as the identifying criteria.
Our results of rapidity ($y$) and transverse momentum ($p_T$) single differential distributions
as well as their double differential distributions of the two states confirm that they are also
valuable criteria.

\section{Brief introduction to PACIAE 3.0 }
The PACIAE model~\cite{sa2012} is a phenomenological model
simulating relativistic elementary particle collisions and nuclear collisions based on
the PYTHIA 6.4 code~\cite{sjostrand2006}. It has been
successfully used for describing the hadron yields, the transverse momentum and rapidity
distributions, the strangeness enhancement, the nuclear modification factor,
and the flow asymmetry, etc. in the relativistic elementary particle collisions and nuclear
collisions~\cite{sa2012,sa2014,zheng2018,lei2023}. The latest version of PACIAE 3.0~\cite{lei2023} is employed in this work.

In the PACIAE 3.0 C-simulation framework each nucleon-nucleon (NN)
or hadron-hadron ($hh$) collision is executed by PYTHIA~\cite{sjostrand2006}
with presetting the hadronization turning-off
temporarily. Therefore an initial partonic state is available after QCD hard
scattering and associated initial- and final-state QCD radiations. A process of
the gluons breaking-up and energetic quarks and antiquarks [(anti)quarks in
short] deexcitation is then executed. In the followed partonic rescattering stage
the lowest-order perturbative quantum chromodynamics (LO-pQCD) parton-parton
interaction cross section~\cite{combridge} is employed. The resulting partonic
final state [PFS, quark phase space (QPS), or quark matter (QM)] comprises
abundant (anti)quarks with their four coordinates and four momenta.

\begin{center}
\begin{figure*}[htbp]
\centering
\includegraphics[width=0.65\textwidth]{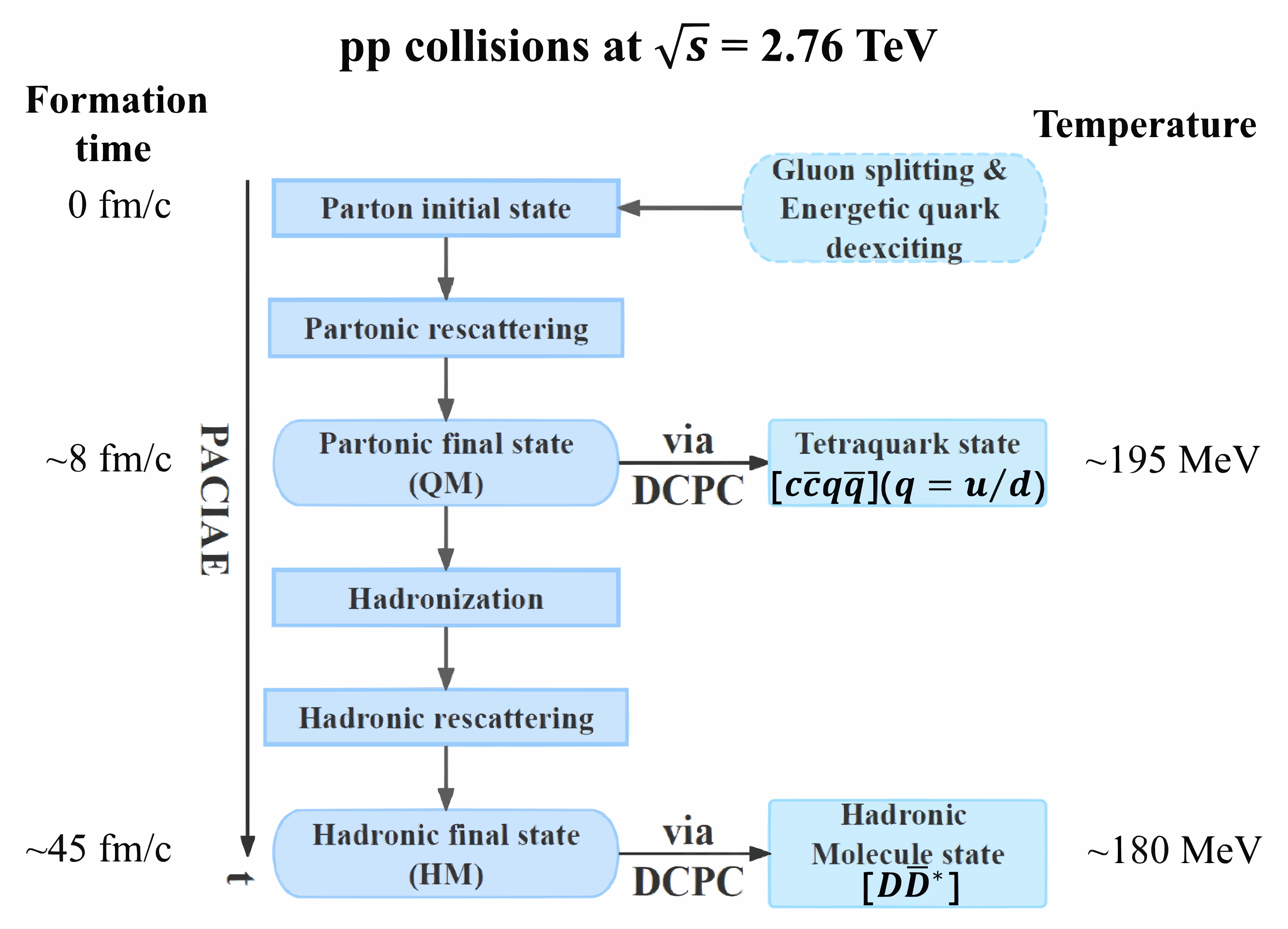}
\caption{A sketch of the $\rm X(3872)$ tetraquark state and molecular state
productions in $pp$ collisions at $\sqrt{s}= 2.76 \,\mathrm {TeV}$ in the
{\footnotesize PACIAE + DCPC} model.}
\label{sket1}
\end{figure*}
\end{center}

After partonic rescattering, the hadronization is implemented by the Lund
string fragmentation regime and/or the coalescence model~\cite{lei2023}
generating an intermediate hadronic state. It is followed by the hadronic
rescattering with result of a hadronic final state (HFS, hadron phase space
(HPS), or hadron matter (HM)) for a NN ($hh$) collision system. This final
hadronic state is composed of numerous hadrons with their four coordinates and
four momenta. The schematic structure of above transport processes is shown by
the left part of the block diagram in Fig.~\ref{sket1}.

\section{DCPC model}

In the quantum statistical mechanics~\cite{stowe2007,kubo1965},
the yield of the N-particle (the partons in PFS or the hadrons in HFS) cluster reads
\begin{eqnarray}
Y_{N}=\int\cdots\int_{E_\alpha\le E\le E_\beta}\frac{d\vec{q}_{1}d\vec{p}_{1}\cdots d\vec{q}_{N}d\vec{p}_{N}}{h^{3N}},
\label{eq: two}
\end{eqnarray}
where the $E_{\alpha}$ and $E_{\beta }$ are the particle lower and upper
energy thresholds. The $\vec {q}_{i}$ and $\vec {p}_{i}$ are, respectively, the
$i$th particle three coordinates and three momenta.
If the cluster exists naturally, certain constraints (the component constraint,
the coordinate constraint, and the momentum constraint) should be satisfied.
For instance, the yield of the $c\overline c u \overline u$ cluster is calculated by

\begin{eqnarray}
Y_{\rm cluster}=\int \dots \int {\delta_{1234}
	\frac{\prod_{i=1}^{i=4} d\vec{q}_{i}d\vec{p}_{i}}{h^{12}}}
\label{eq: three},
\end{eqnarray}
where
\begin{eqnarray}
\delta_{1234}=\left\{\begin{array}{ll}
	1  \mbox { if } [1 \equiv c,  2 \equiv \overline c,
	                 3\equiv u, 4 \equiv \overline u , \\
\quad m_{\rm 0}-\Delta m \leq m_{inv}\leq m_{\rm 0}+\Delta m, \\
         \hspace{0.25cm} |\vec{q}_{ij}| \leq R_{0} \hspace{0.25cm}
	(i\neq j;i,j=1,2,3,4)]; \\[4pt]
    0  \mbox { otherwise. }
\end{array}\right.
\label{eq: three_1}
\end{eqnarray}

In Eq.~(\ref{eq: three_1}), $m_0$ denotes the cluster mass and $\Delta m$ is
the mass uncertainty (a parameter). The invariant mass, $m_{\rm inv}$,
reads
\begin{eqnarray}
m_{\rm inv}=\sqrt{\bigg(\sum^{4}_{i=1} E_i \bigg)^2-\bigg(\sum^{4}_{i=1}
\vec p_i \bigg)^2},
\label{eq:two_3}
\end{eqnarray}
where $E_i$ and $\vec p_i$ ($i$=1,2,3,4) are, respectively, the energy and
three-momentum of the component particles ($c,\overline c, u, \overline u$).
The $R_0$ and $|\vec{q}_{ij}|$ denote, respectively, the radius of the cluster
(a free parameter) and the relative distances between any two component
particles.

The $\rm X(3872)$ tetraquark state is coalesced (hadronized) in QPS
by the DCPC model with four component particles of $c$ and $q$
(anti)quarks ($ q=u/d $, the same later). To the end, we first construct a
component particle list based on the (anti)quark list in FPS (i.e., QPS). A
four-layer cycle over component particles in the list is then built. Each
combination in this cycle, if it is composed of $c$ and $q$ (anti)quarks and
satisfied the constraints of Eq.~(\ref{eq: three_1}), counts as an
$\rm X(3872)$ tetraquark state. The component particle list is then updated by
removing the used (anti)quarks, resulting in a new component particle list. A
new four-layer cycle is built, etc. Repeat this process until the emptying of
the component particle list or the rest of the list is unable to form an
$\rm X(3872)$ tetraquark state. The parameters in Eq.~(\ref{eq: three_1})
are set as follows: $m_0=m_{\rm X(3872)}= 3666
\,\mathrm{MeV/c^{2}}$ (sum of the masses of constituent
(anti)quarks~\cite{pdg2018}), $\Delta m=1.95\,\mathrm{MeV/c^{2}}$ (estimated
by the half decay width of $\rm X(3872)$~\cite{ge2021,belle2010}), and
$R_{0}<1.0\,\mathrm{fm}$ (as the quark is a pointlike
particle~\cite{bao2022,grinstein2024}).

Similarly, the $\rm X(3872)$ molecular state is recombined with eight component
mesons of $D^+$ and $D^-$, $D^0$ and $\bar{D}^0$ and their excited states in
HPS (i.e., HFS) by the {\footnotesize DCPC} model. We first construct a
component meson list based on the hadron list in HFS. A two-layer cycle over
component mesons in the list is then built. Each combination in the cycle,
if it is zero charge and satisfies the corresponding constraints similar to
Eq.~(\ref{eq: three_1}), counts as an $\rm X(3872)$ molecular state. The
parameters in the above equation are set as follows:
$m_0=m_{\rm X(3872)}= 3872\,\mathrm{MeV/c^{2}}$ (identifying the experimentally
observed peak of $\rm X(3872)$ as the molecular state) and
$\Delta m=142\,\mathrm{MeV/c^{2}}$ (obtained from
$2m_{D} < \Delta m < 2m_{D^{*}}$~\cite{hui2021}). The $R_{0}$ is in the range
of $1.0\,\mathrm{fm} < R_{0} < 10.0\,\mathrm{fm}$~\cite{bao2022,grinstein2024}
($R_{0}$ should be larger than the sum of the radius of two component mesons
and less than the interaction range of $20\,\mathrm{fm}$).

\section{Apparent temperature prediction}

We follow Ref.~\cite{rosales2024} introducing the Shannon
entropy~\cite{rosales2024,shannon1948} of
\begin{equation}
 \mathcal{H}=-\int_{0}^{\infty}(\mathrm{TMD}/I_{0})\ln[\mathrm{TMD}/I_{0}]dp_{T},
\label{ther1}
\end{equation}	
where $\mathrm {TMD}$ and $I_{0}$ refer to the particle's transverse momentum
distribution and its first moment, respectively. If the considered particle's
transverse momentum distribution can be fitted by a Hagedorn-like $p_T$
distribution
\begin{equation}
\frac{dN}{dp^{2}_{T}}\propto (\frac{p_{0}}{p_{T}+p_{0}})^m,
\label{ther2}
\end{equation}
Eq.~(\ref{ther1}) reads
\begin{equation}
\mathcal{H}=\frac{m}{m-1}+\ln(\frac{m}{m-1})+\ln(T).
\label{ther3}
\end{equation}	
The parameters in the above equations are
\begin{eqnarray}
m(\sqrt s)=a_m(\frac{s}{s_0})^{c_m/2}, \\
p_0(\sqrt s)=a_{p_0}(\frac{s}{s_0})^{c_{p_0}/2},
\end{eqnarray}
with $a_m=8.45$, $c_m=-0.082$, $a_{p_0}=1.22\,\mathrm {GeV}$, and $c_{p_0}=-0.03\,\mathrm {GeV}$.

We first calculate the Shannon entropy of HM and QM by Eq.~(\ref{ther1}) with
the simulated $\pi^{+}+\pi^{-}$ $p_T$ distribution in HM and
$u+\bar d+d+\bar u$ $p_T$ distribution in QM, respectively. The HM and QM
Shannon entropies are then inserted, respectively, into Eq.~(\ref{ther3}),
resulting in $T_{\rm HM}\equiv T_{\rm molec}=180\,\mathrm {MeV}$ and $T_{\rm QM}\equiv T_{\rm tetra}=195\,\mathrm {MeV}$,
which are independent of the DCPC model.

\section{Results and discussion}

The main model parameters of the allowed number of deexcitation generation and
the threshold energy of deexcitation in the energetic (anti)quark deexcitation
process, i.e., the adj1(16)=1 and adj1(17)=1.8 $\mathrm{GeV}$, are fixed by
fitting the ALICE data of $D^0$ and $D^+$ yield in $pp$ collisions at
$\sqrt{s}=2.76\,\mathrm{TeV}$~\cite{alice20127}. The partonic final state
(QPS, QM) and hadronic final state (HPS, HM) are then simulated by the
{\footnotesize PACIAE 3.0} model with total of two billion $pp$ collision
events. Subsequently, the $\rm X(3872)$ tetraquark state and molecular
state are coalesced and recombined, respectively, in the QM and HM by
{\footnotesize DCPC} model.

The formation time of the $\rm X(3872)$ tetraquark state (i.e., the
average evolution time upto QM) or of the molecular state (i.e., the average
evolution time upto HM) is calculated by
\begin{eqnarray}
\langle t \rangle= \frac{\sum^{N_{\rm eve}}_{i=1} t_{i} }{N_{\rm eve}},
\label{eq:ft}
\end{eqnarray}
where $t_{i}$ stands for the evolution time upto QM or HM in the $i$th event, and
$N_{\rm eve}$ is the total number of simulated events. The calculated result is
$\langle t \rangle_{\rm tetra}=8\,\mathrm {fm/c}$ or
$\langle t \rangle _{\rm molec}=45\,\mathrm {fm/c}$.

Following the conventional Monte Carlo coalescence model~\cite{lei2023,linzw2005}
we assume the four momenta of the tetraquark (molecular) state is the sum of the four momenta
of its component particles. Thus the velocity of an $\rm X(3872)$
tetraquark state composed of $c$, $\overline c$, $u$, and $\overline u$ is
calculated by
\begin{eqnarray}
 \vec v= \frac {\sum^{4}_{i=1}\vec p_i} {\sum^{4}_{i=1} E_{i}},
\label{eq:v1}
\end{eqnarray}
for instance. The corresponding average velocity $\langle v \rangle$ reads
\begin{eqnarray}
\langle v \rangle= \frac {\sum^{N}_{j=1} v_{j}}{N},
\label{eq:v2}
\end{eqnarray}
where $v_{j}$ refers to the velocity of the $j$th tetraquark state and $N$ stands
for the total number of tetraquark states. The same is for the calculation of
average velocity of molecular states. Thus we have the result of
$\langle v \rangle_{\rm tetra}=0.96 c$ or $\langle v \rangle_{\rm molec}=0.73 c$.

A largish discrepancy between the two $\rm X(3872)$ states in the formation
time, average velocity, and temperature is observed. Each of them is then
proposed as a criterion (probe) distinguishing the $\rm X(3872)$ tetraquark
state or molecular state. We note that the discrepancy
between the $\rm X(3872)$ multiquark state and molecular state in the formation
time, average velocity, and temperature increases with increasing collision
energy and size of collision system.

\begin{center}
\begin{figure*}[htbp]
\centering
\includegraphics[width=0.7\textwidth]{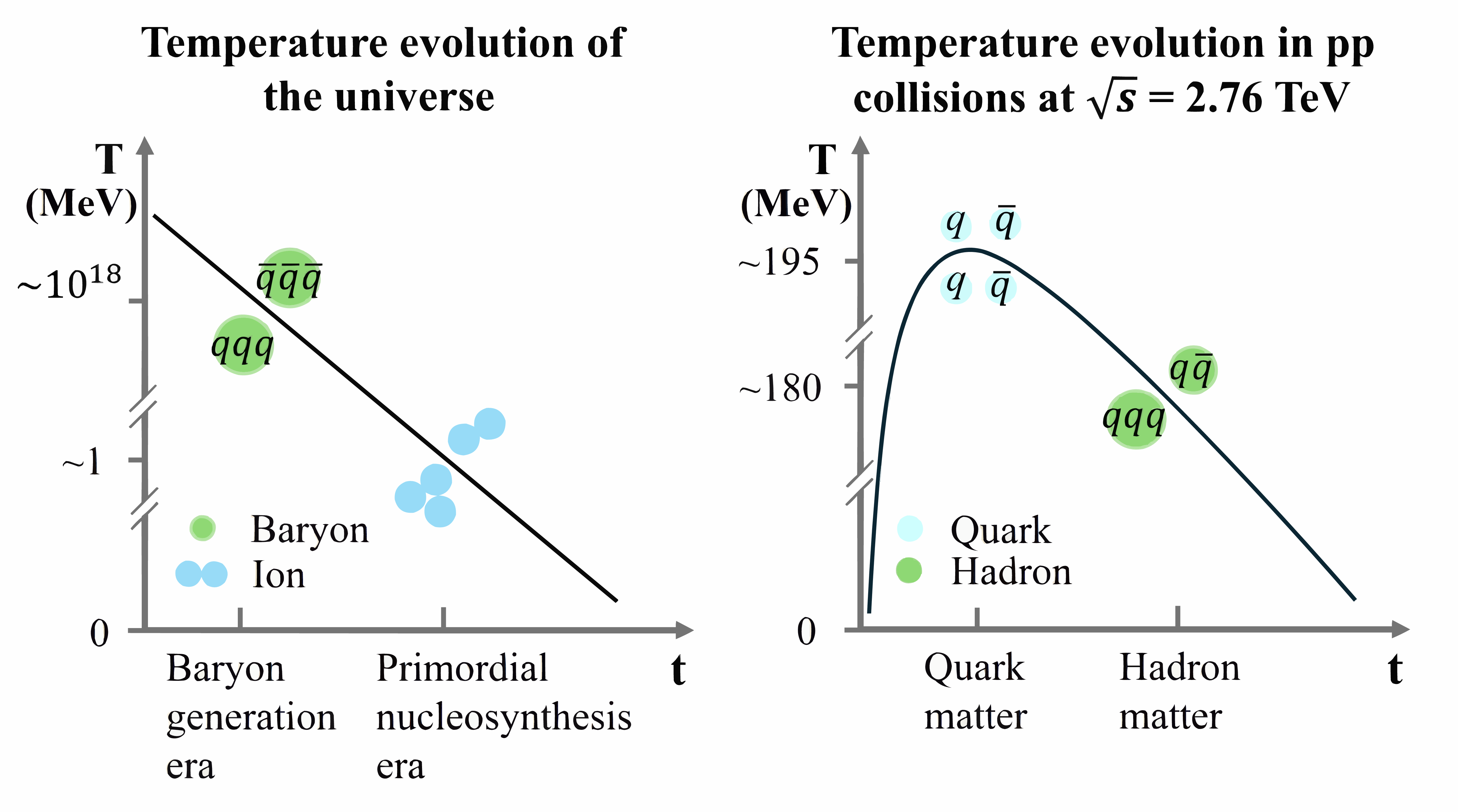}
\caption{A sketch for the temperature evolution process in the universe and
	 the $pp$ collisions at $\sqrt{s}=2.76\,\mathrm {TeV}$, respectively.}
\label{tem}
\end{figure*}
\end{center}

The QGP phase transition in the relativistic heavy ion collisions (HICs) and/or
$pp$ collisions is highlighted as a reproduction of the big bang in the
universe~\cite{yang1981} originally. One step further, Fig.~\ref{tem} gives a
qualitative comparison between the temperature evolution process in the early
universe~\cite{david1984} and in the $pp$ collisions at $\sqrt{s}=2.76\,
\mathrm {TeV}$. The QM and HM two main stages in the evolution of relativistic
$pp$ (HIC) collisions are assumed to be corresponding to the
baryon generation and the primordial nucleosynthesis two major
eras~\cite{david1984} in the universe, respectively. Hopefully, more related investigations
would be promoted.

\begin{figure*}[htbp]
\includegraphics[scale=0.85]{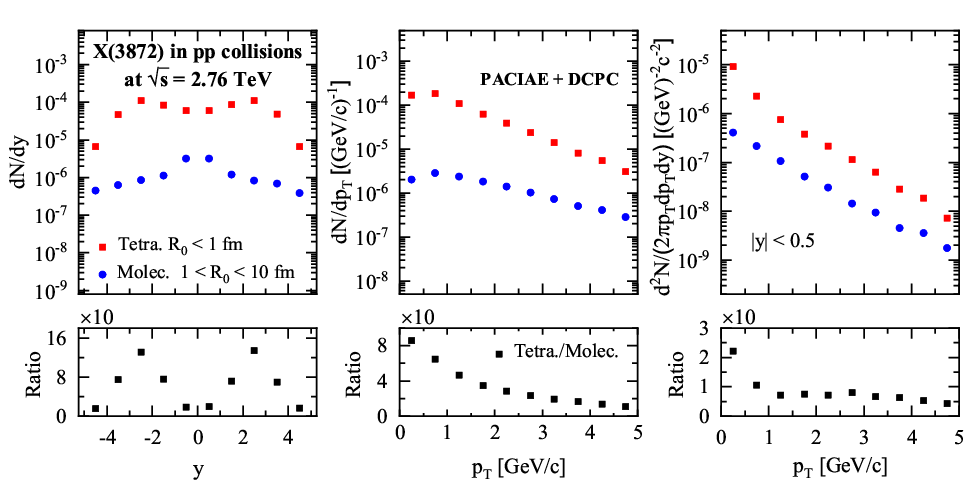}
\caption{The simulated $y$ and $p_T$ single differential distributions as well
as $p_T$ and $y$ double differential distributions of the $\rm X(3872)$ two states
in $pp$ collisions $\sqrt{s}=2.76\,\mathrm {TeV}$.}
\label{common}
\end{figure*}

Meanwhile, the single rapidity ($y$) and $p_T$ differential distributions, as
well as the $p_T$ and $y$ double differential distributions of the $\rm X(3872)$
two states are also calculated and shown in Fig.~\ref{common}. A significant
discrepancy between the two $\rm X(3872)$ states is observed in all of those
distributions. We confirm that each of them is also an effective criterion
distinguishing the $\rm X(3872)$ two states.

In the PACIAE simulation, there is a QM stage, HM stage,
and a process from QM to the HM. It means that the QGP phase transition exists
in the simulation process. If the above proposed criteria (probes) are employed
distinguishing the $\rm X(3872)$ two states successfully in any direct and/or
indirect experimental measurements, it will be exciting. It proves, indeed, the
QGP phase transition. Investigating these criteria (probes) plays a role in the
different energy and/or size of collision systems, which is also a study for the QGP
phase transition in different collision systems.

The study should be extended to $pp$ collisions at $\sqrt{s}= 5.02$ and $13.6\,\mathrm {TeV}$
as well as Pb-Pb collisions at LHC energies. A discrepancy between
the $\rm X(3872)$ tetraquark state and molecular state in central and forward
rapidity regions as well as in central and peripheral Pb-Pb collisions
deserves to be studied further. On the other hand, the nuclear modification
factor ($R_{AA}$) might also be good identifying criterion and worthy to be
investigated.

\section*{Acknowledgements}
The authors thank Shi-Lin Zhu, Xiao-Ming Zhang, Li-Lin Zhu, and Meng-Zhen Wang for helpful discussions.
This work was supported by the National Natural Science
Foundation of China under Grant No. 12375135 and by the
111 Project of the Foreign Expert Bureau of China. Y.L.Y. acknowledges the
financial support from the Key Laboratory of Quark and Lepton Physics of the Central
China Normal University under Grant No. QLPL201805 and the Continuous Basic
Scientific Research Project (Grant No. WDJC-2019-13). The work of W.C.Z. is supported
by the Natural Science Basic Research Plan in Shaanxi Province of China(Program No. 2023-JCYB-012).


%

\end{document}